\documentstyle[aps,psfig]{revtex}
\begin{document}
\title{
\hfill{\small {\bf MKPH-T-03-6}}\\
{\bf Do we understand the $\eta N$ interaction from the near threshold 
$\eta$ photoproduction on the deuteron?}}
\author{A.\,Fix
\footnote{On leave of 
absence from Tomsk Polytechnic University, 
634034 Tomsk, Russia} 
and H.\,Arenh\"ovel}
\address{Institut f\"ur Kernphysik,
Johannes Gutenberg-Universit\"at Mainz, D-55099 Mainz, Germany}
\date{\today}
\maketitle
\begin{abstract}
The effects of final state interaction in incoherent $\eta$ photoproduction 
on deuteron are studied within a three-body approach including a realistic 
$NN$ potential. The results are compared
with available data, and differences with other theoretical predictions are 
analyzed. The role of the $\eta N$ interaction and the possibility of 
extracting the $\eta N$ scattering parameters from this reaction are 
discussed. 
\end{abstract}



\section{Introduction}\label{sect1}

The role of final state interaction (FSI) in incoherent photoproduction of 
$\eta$-mesons on the deuteron was investigated in our previous work 
\cite{FiAr97,FiAr02}. First we have considered in \cite{FiAr97} 
the first-order approximation (FOA) where complete hadronic rescattering is 
included in the two-body subsystems ($NN$ and $\eta N$) only,  
and in the subsequent work \cite{FiAr02} the three-body aspects of the 
reaction were studied. Summarizing the 
results we would like to conclude that the importance of final 
state interaction arises from two sources which are interrelated. 
Firstly, the impulse approximation (IA), where the FSI is ignored, predicts 
a very strong suppression of the 
cross section close to the threshold. The reason for this is a
strong mismatch between the large nucleon momentum needed for 
$\eta$ meson production (more than 200 MeV/c)
and the average available momentum in the deuteron of about 45 MeV/c. The 
FSI provides an efficient 
mechanism to compensate this momentum mismatch 
by the collision between the final particles. As a 
consequence, already the leading terms in the multiple scattering 
series associated with pairwise $NN$ and $\eta N$ scattering produce a very 
large enhancement of the $\eta$ production rate close to the threshold. 
This conclusion has later been confirmed by Sibirtsev et al.\ 
\cite{Sib01,Sib02b,Sib02c}. 

The second reason is a 
relatively strong $\eta N$ interaction at low-energy. 
Because of its amplification via the 
$NN$ interaction an appreciable attraction in the 
$\eta NN$ system is generated, which in turn 
leads to a virtual pole in the $S$-wave part of the three-body 
scattering amplitude \cite{Deloff00,FiAr00,Wycech01}. 
In the case of the spin-singlet $\eta NN$ state  
$(J^\pi,T)=(0^-,1)$, which dominates the reaction 
close to the threshold, the pole lies on the three-body unphysical 
sheet not far from 
zero kinetic energy \cite{FiAr00}. We would like to 
emphasize that the existence of a virtual 
state near zero energy is the reason, why the FOA 
cannot provide an accurate description of the 
reaction dynamics at low energy. This statement is corroborated by the fact, 
that the multiple scattering series  
converges slowly near the pole so that the FOA 
does not constitute a reliable approximation to the whole series. 
In particular, the FOA is unable to account for anomalies in the energy
dependence of the total cross section, since singularities of the 
three-body scattering amplitude cannot be generated in a perturbative 
approach (see, e.g., the general arguments given in \cite{Goeb64}). 
Indeed, as was shown in \cite{FiAr02}, only the 
full three-body treatment can explain, at least qualitatively, the  
anomalously strong rise of the experimental $\gamma d\to \eta X$ cross 
section just above threshold. 

However, our work in \cite{FiAr02} was only considered as a first step for  
taking into account the most important dynamical properties of the 
$\eta NN$ system, because of several simplifications for the two-body 
forces. This concerns primarily the $NN$-sector, 
especially the deuteron wave function, which in \cite{FiAr02} was  
taken as a pure $^3S_1$ state, generated by 
the Yamaguchi potential \cite{Yam54}. 
Clearly, such a treatment is too simple for $\eta$
photoproduction where large momentum transfers come into play.

Thus the first motivation for the 
present paper is to overcome this shortcoming by using a realistic
$NN$-potential. In Sect.~\ref{sect2}, we give the details of the two-body 
interactions and present the results for the total and differential 
cross sections. For the comparison with the inclusive data of \cite{Hej02} 
also the coherent reaction $\gamma d\to\eta d$ is calculated. As will be 
shown, our predictions slightly 
underestimate the data in almost the whole energy 
region from threshold up to $E_\gamma = 750$ MeV. 
We will discuss possible reasons for this disagreement. 
The second motivation of the paper is to compare our results with the  
work of Sibirtsev et al.~\cite{Sib01,Sib02b,Sib02c}. 
In Sect.~\ref{sect3} we point out principal disagreements  
between the results of \cite{Sib01,Sib02b,Sib02c} and ours, 
which cannot be explained by the differences of the model ingredients. 
Finally, in Sect.~\ref{sect4} we analyze the possibility of 
extracting the strength of the $\eta N$ interaction in 
incoherent $\eta$ photoproduction on the deuteron. 


\section{Two-body input and discussion of the results}\label{sect2}

The transition matrix element of the reaction $\gamma d\to\eta np$ has been 
evaluated with inclusion of the hadronic interactions between the final 
particles whereas the initial electromagnetic interaction is  
treated perturbatively in lowest order. The general formalism of our 
approach is described in detail in \cite{FiAr02}. 
Here we present mainly the most important two-body ingredients of the 
calculation.

As a basic input we need the $\eta N$ and $NN$ scattering amplitudes which
were restricted to $S$-states only in view of the near threshold region. 
For the $\eta N$ interaction we use a conventional isobar ansatz as 
described in detail in \cite{BeTa91}, where  
the $\eta N$ channel is coupled with $\pi N$ and $\pi\pi N$ channels 
through the excitation of the $S_{11}(1535)$ resonance. 
The separable $\eta N$ scattering matrix has the usual isobar form 
\begin{equation}\label{eq10}
t_{\eta N}(q,q';W)=\frac{g_\eta(q)g_\eta(q')}
{W-M_0-\Sigma_\eta(W)-\Sigma_\pi(W)+\frac{i}{2}\Gamma_{\pi\pi}(W)}\,,
\end{equation}
as a function of the invariant energy $W$. 
The $S_{11}(1535)$ self energies 
$\Sigma_\eta$ and $\Sigma_\pi$ are  
determined by the vertex functions $g_\alpha(q)$ as 
\begin{equation}\label{eq20}
\Sigma_\alpha(W)=\frac{1}{(2\pi)^2}\int\limits_0^\infty 
\frac{q^2\,dq}{2\omega_\alpha}
\frac{[g_\alpha(q)]^2}
{W-E_N(q)-\omega_\alpha(q)+i\epsilon}\,, \  \quad \alpha\in\{\pi,\eta\}\,,
\end{equation}
with $E_N$ and $\omega_\alpha$, ($\alpha\in\{\pi,\eta\}$)
denoting the on-shell 
energies of nucleon and meson, respectively.  
The two-pion channel is included in a simplified manner 
by adding the $S_{11}\to\pi\pi N$ 
decay width, parametrized by   
\begin{equation}\label{eq30}
\Gamma_{\pi\pi}(W)=\gamma_{\pi\pi}\frac{W-M_N-2m_\pi}{m_\pi}\,, \quad
\mbox{with} \quad \gamma_{\pi\pi}=4.3\ \mbox{MeV}\,.
\end{equation}
The vertex functions are taken in a Hulth\'en form 
\begin{equation}\label{eq40}
g_\alpha(q)=g_\alpha\bigg[1+\frac{q^2}{\beta_\alpha^2}\bigg]^{-1}\,, 
\end{equation}
containing the strength of the coupling $g_\alpha$ and the range of 
the Hulth\'en form factor $\beta_\alpha$. 
The parameters (see Table~\ref{tab1}) 
were adjusted to fit the $\eta N$ scattering length  
\begin{equation}\label{eq50}
a_{\eta N}=(0.5+i0.32)\ \mbox{fm}\,,
\end{equation}
and at the same time to provide a reasonably good description
of the reactions $\pi N\to\pi N$ and $\pi N\to\eta N$ (for more details
see \cite{FiAr03}). We consider this value as an approximate
average of the scattering lengths provided by modern $\eta N$ scattering 
analyses (see, e.g., the compilation in Table\,I of \cite{Sib02b}). 
In Sect.~\ref{sect4}, where we study the 
dependence of the results on the $\eta N$ interaction strength, 
two other sets of $\eta N$ parameters are used.
The first one, giving $a_{\eta N}=(0.25+i0.16)$ fm is taken from 
\cite{BeTa91}. The second set is adjusted such, that the scattering length 
$a_{\eta N}=(0.75+i0.27)$ fm of the analysis \cite{Wyc97} is reproduced. 
The corresponding parameters are listed in Table~\ref{tab1}. 

We would like to emphasize that the latter value 
is obtained simply by varying the parameters of our 
separable ansatz (\ref{eq10}) without using the original model of \cite{Wyc97}.  
One of the consequences of this strategy is that the Born term introduced in 
\cite{Wyc97} is absorbed in our approach by the resonance amplitude, which 
leads to an overestimation of the $S_{11}$ contribution. 
In this case we achieve a satisfactory description 
of the $\pi N\to\eta N$ experimental cross section, and the resulting $\eta N$
amplitude agrees rather well with the ones of \cite{Wyc97}. 
However, unlike the 
first two sets, we cannot reproduce the $S_{11}$ wave of the 
$\pi N\to\pi N$ analysis \cite{VIP}, also below the $\eta N$ threshold,  
unless a relatively large contribution of the $S_{11}(1650)$ resonance 
is included. 

The electromagnetic vertex of the amplitude 
$t_{\gamma p}$ for the elementary process $\gamma p\to\eta p$
was fixed by fitting the corresponding data of \cite{Kru95}. 
Below the $\eta N$ threshold, we have required that the pion production 
amplitude $\gamma p\to S_{11}(1535)\to\pi N$, taken from \cite{MAID}, is 
reasonably well described as presented in \cite{FiAr03}. 
For the neutron amplitude we have used the relation 
\begin{equation}\label{neut}
t_{\gamma n} = -0.82\, t_{\gamma p}\,,
\end{equation}
which is consistent with the experimental value 
$\sigma_{\gamma n}$ = 0.67 $\sigma_{\gamma p}$
extracted from $\eta$ photoproduction on very light nuclei 
\cite{Hoff97,Weiss03}.

With respect to the $NN$-interaction, only the most important $^1S_0$ state 
is taken into account, since, as was shown in \cite{FiAr97}, 
the contribution of the triplet state is insignificant, primarily due to the  
isovector nature of the electromagnetic excitation $\gamma N\to S_{11}$. 
For the $^1S_0$ state we have used the separable representation BEST3
\cite{Hei86} for the Bonn potential. The deuteron bound state 
wave function was calculated using the corresponding 
BEST4 version for the $^3S_1-^3D_1$ states. 

The three-body integral equations were solved only for the lowest 
$S$-wave three-body configuration 
where the orbital angular momentum $l=0$ in the two-body 
subsystems is coupled with angular momentum $l=0$ of the third 
particle with respect to the pair. The remaining partial waves were treated 
perturbatively up to the first order in the $S$-wave $t$-matrices of 
$NN$ and $\eta N$ scattering. This approximation is well justified by 
the strong $S$-wave dominance in the 
$NN$ and $\eta N$ low energy interactions. As was shown in 
\cite{FiAr02}, it is the lowest $S$-wave three-body state, that 
is very sensitive to the higher-order scattering contributions, whereas 
the higher partial waves are well approximated by the first order terms. 
As already indicated above, perturbative calculation, where only pairwise 
$NN$ and $\eta N$ FSI's are taken into account in all partial waves 
is referred to as first-order approximation (FOA).

\begin{figure}
\centerline{\psfig{figure=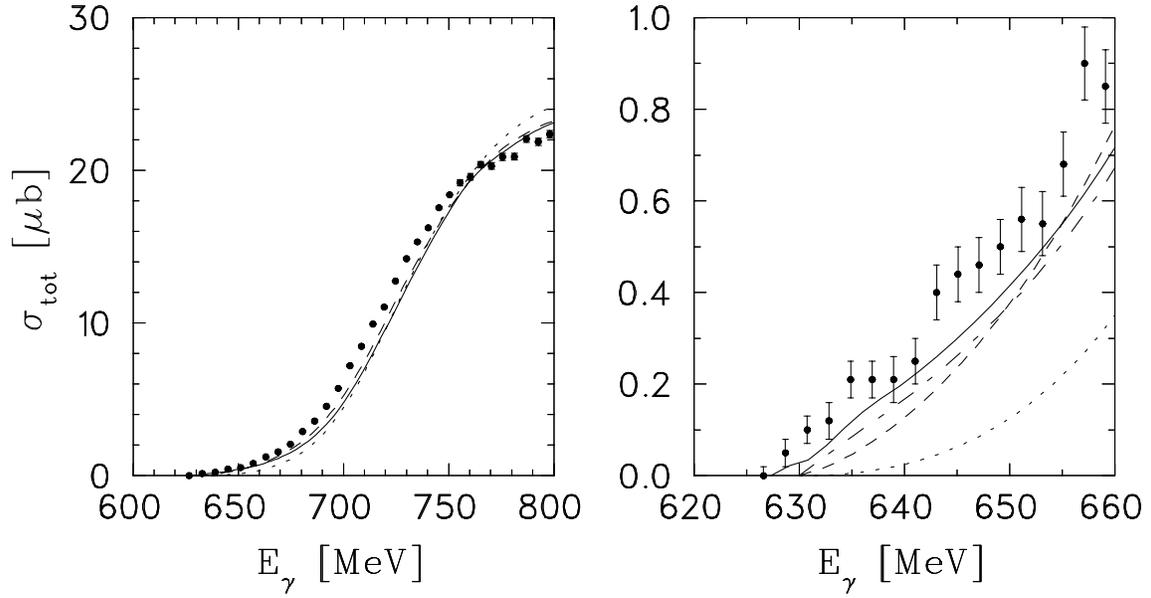,width=15cm,angle=0}}
\vspace{.5cm} 
\caption{
Total cross section for the reaction $\gamma d\to\eta np$. The near-threshold
region is shown separately in the right panel. The dotted, dashed and 
dash-dotted lines correspond to the impulse approximation 
(IA), first-order calculation (FOA) and three-body model for the 
final $\eta NN$ state. The inclusive cross section $\gamma d\to\eta X$, 
obtained by adding the 
coherent cross section (see Fig.\,\protect\ref{fig3}) is shown by the solid 
curve. The inclusive data are taken from \protect\cite{Hej02}. 
The dash-dotted and solid curves are indistinguishable in the left panel. 
} 
\label{fig1}
\end{figure}

\begin{figure}
\centerline{\psfig{figure=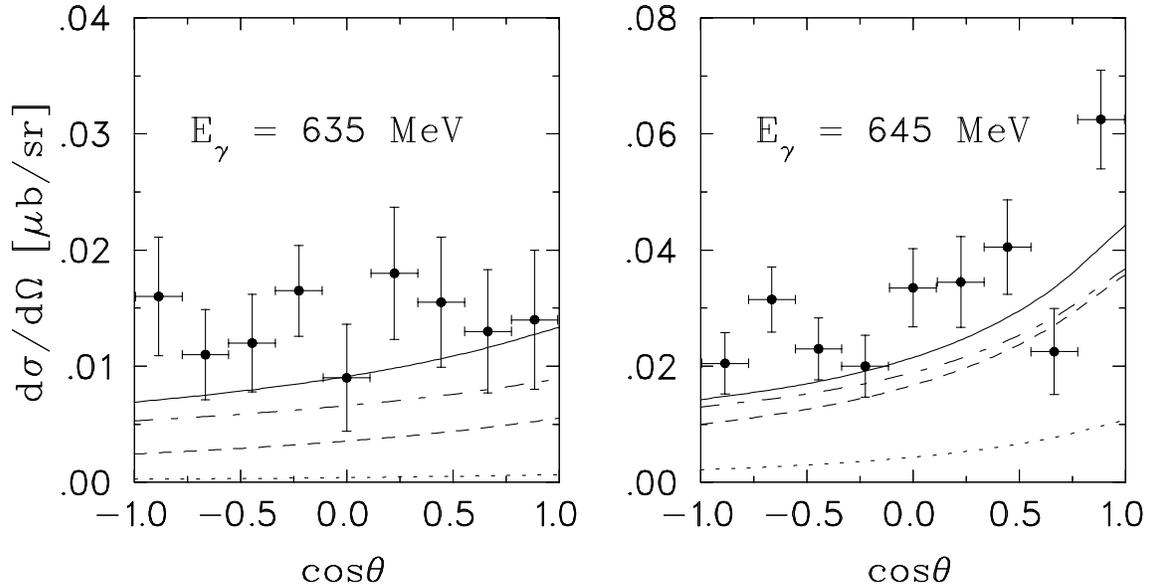,width=15cm,angle=0}}
\vspace{.5cm} 
\caption{
Differential cross section for  $\gamma d\to\eta np$. 
Notation as in Fig.\,\protect\ref{fig1}. 
} 
\label{fig2}
\end{figure}

The significance of FSI is demonstrated in Fig.\,\ref{fig1} for 
the total cross section and in Fig.\,\ref{fig2} for the $\eta$ 
angular distribution. In order to appreciate the importance of a realistic 
treatment of the $NN$ sector, 
the results should be compared with those presented in 
Figs.\,8 and 9 of \cite{FiAr02}, obtained by means of the Yamaguchi potential
and using a pure $S$-wave deuteron. 
As expected, for a realistic $NN$-interaction  the FSI effect becomes 
smaller. The obvious reason is the $NN$-repulsive core which weakens the 
attraction in the final $\eta NN$ system at small relative distances. At 
the same time the ``three-body'' effect remains important. 
For instance, at $E_\gamma = 635$ MeV the three-body treatment enhances the 
first-order result by about a factor two. Also the steep rise right above 
threshold is a characteristic feature of the three-body approach which is
not born out in FOA. 

\begin{figure}
\centerline{\psfig{figure=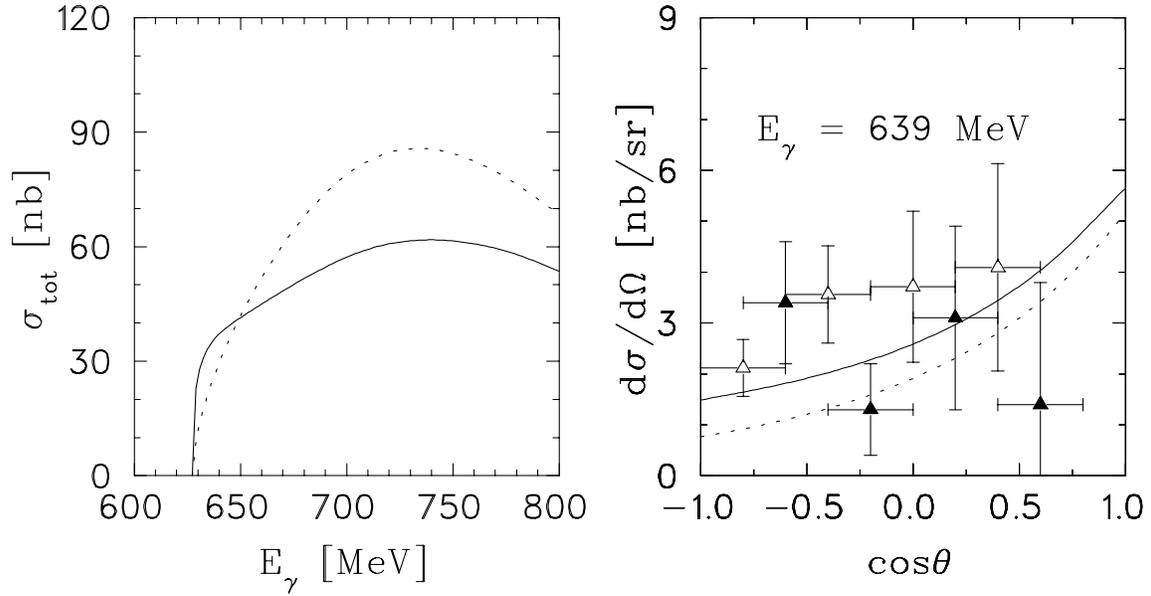,width=15cm,angle=0}}
\vspace{.5cm} 
\caption{
Total and differential cross sections for coherent eta 
photoproduction on the deuteron. The IA and three-body results are shown 
by the dotted and the solid curves, respectively.  
The data are from \protect\cite{Hoff97} (open triangles) and 
\protect\cite{Weiss01} (filled triangles).} 
\label{fig3}
\end{figure}

In order to compare our results with the inclusive measurement 
\cite{Hej02} we calculated in addition to the break-up channel 
the coherent cross section 
$\gamma d\to\eta d$.
In analogy with the incoherent process, 
the calculation was performed within the quasiparticle formalism
of the three-body problem, as described in \cite{FiAr02}.   
In this case we neglect the small contribution of the deuteron $D$-state, 
which contributes only 1$\%$ in the IA cross section. 
Since the coherent cross section is proportional to the 
modulus of the isoscalar part $g_{(s)}$ of the $\gamma N\to S_{11}$ 
transition amplitude, we have fixed it according to the relation 
$\alpha=|g_{(s)}|/|g_{\gamma p\to S_{11}}|=0.25$ 
as found in \cite{RiAr01}. 
This value is also close to $\alpha=0.22$ of \cite{BeTa91}. 
The corresponding total cross section and one   
angular distribution are plotted in Fig.\,\ref{fig3}.  
As one notices, the effect of FSI is also very pronounced. 
It is not, however,  
as large as obtained in \cite{FiAr02}, where we have used a more attractive  
$\eta N$ interaction with $a_{\eta N}=(0.75+i0.27)$ fm. 

Adding the coherent contribution we obtain the inclusive cross section 
shown in Figs.~\ref{fig1} and \ref{fig2} by solid curves. 
We notice that although the three-body calculation leads to a sizable
improvement of the theoretical prediction just above threshold, 
a quantitative agreement with the experimental results has not been reached 
yet, the theory being too low. 
Moreover, above  $E_\gamma = 650$ MeV the inclusion of 
higher-order terms in the lowest partial wave acts in the opposite 
direction by decreasing the cross section of the FOA. 
In this region the three-body model exhibits an even larger 
deviation from the data than the FOA.

The slight 
disagreement with the experimental results points apparently to the fact, 
that the mechanism of the $\eta$ photoproduction is more complicated and 
some of its 
important details are not properly accounted for by our calculation. 
In this connection, we would like to make a few comments concerning 
possible ways of improving the theoretical treatment.
The first relates to the 
off-shell behavior of the $\eta N$ scattering matrix, which may be much more 
complex as given by the vertices $g_\eta(q)$ in (\ref{eq40}). 
In particular, it was already noted in \cite{FaWi01} that the simple Hulth\'en 
form factors may strongly overestimate the short-range 
$\eta NN$ interaction, since the resulting separable $\eta N$ potential 
is probably too attractive near the 
origin. This is hardly important for low-energy $\eta d$ scattering 
but it is relevant for $\eta$ photoproduction 
which in general is quite sensitive to the  
$\eta N$ wave function at small distances. Clearly, 
a more realistic description of this short-range behavior would require 
a much more thorough treatment of the $\eta N$ interaction, 
where the resonance excitation 
is considered microscopically with respect to the internal dynamics of 
hadrons. 

Another comment is related to the explicit coupling to 
$\pi NN$ states. It was neglected  
in the present calculation (except for virtual $\pi N$ decays 
in the $S_{11}$ propagator), since a correct inclusion of a pion 
would require substantial refinements of our three-body treatment, 
in particular the insertion of a variety of resonances which are excited 
in $\pi N$-collisions. If the $\pi NN$ states are properly 
taken into account, then among other factors  
the $\eta$ photoproduction can proceed according to the two-step scheme 
$\gamma N\to\pi N\to\eta N$, where a pion, being produced 
by the photon, is subsequently rescattered into an $\eta$ by the 
other nucleon. 
According to the results of \cite{FiAr02,RiAr01,GaPe02},
the contribution of the intermediate pion depends strongly on the 
role of large momentum transfers in the reaction mechanism. 
For example, whereas $\pi NN$ states provide only a small fraction of  
the $\eta d$ scattering cross section \cite{FiAr02,GaPe02}, 
they contribute rather sizeably to 
coherent $\eta$ photoproduction on the deuteron \cite{RiAr01}. 
It was shown in \cite{FiAr02}, where only the $S_{11}(1535)$ 
resonance was included into the $\gamma N\to\pi N$ amplitude, 
that the $\gamma d\to\eta np$ 
cross section is insignificantly affected if the 
pion rescattering mechanism 
is considered. However, it may well be that inclusion of other resonances 
into the pion photoproduction amplitude 
can improve the agreement between our calculation and the 
data \cite{Hej02}. 

Finally, it must be noted that the relation (\ref{neut}), fixing the 
neutron amplitude, is only the simplest variant matching the required 
relation between the elementary cross sections. Namely, as was pointed out 
in \cite{RiAr01}, it does not account for a possible relative phase between  
$t_{\gamma p}$ and $t_{\gamma n}$. This phase is predicted, e.g., if the 
corresponding electromagnetic vertices are extracted by fitting different 
isotopic channels of pion photoproduction \cite{BeTa91,RiAr01}. 
Most likely, this fact is insignificant in the region of large photon 
energies, where the amplitudes $t_{\gamma p}$ and $t_{\gamma n}$ are added 
incoherently because of large relative momenta between the particles. At the 
same time, near threshold the interference between $t_{\gamma p}$ and 
$t_{\gamma n}$ increases due to a reduction of the available phase space, 
which 
requires a more sophisticated treatment of the isotopic structure of the 
elementary photoproduction operator.


\section{Comparison with other work}\label{sect3}

Now we turn to the comparison of our results to those obtained by 
Sibirtsev et al. in a series of papers \cite{Sib01,Sib02b,Sib02c} where the 
FSI effects were studied within the first-order approximation.  
First of all, they confirmed our previous conclusion \cite{FiAr97} 
with respect to the fundamental role of the final state
interaction in the near threshold region. 
But moreover, they have claimed that already the 
incoherent reaction in FOA for the final state without inclusion of 
the coherent channel provides a good 
description of the experimental cross section 
for $\gamma d\to\eta X$ \cite{Hej02} with a reasonable value for the 
$\eta N$ scattering length. 
This result is at variance with our own conclusion about 
the necessity of a three-body treatment of the final $\eta NN$ state, 
and the nonnegligible contribution of the coherent reaction. 
These contradicting conclusions seem to be especially surprising  
in view of the fact that the models  
used in \cite{Sib01,Sib02b,Sib02c} 
and in \cite{FiAr97,FiAr02} differ only in nonessential 
details, such as, e.g., relativistic vs.\ nonrelativistic 
parametrizations of the $\eta$ production 
amplitude. Therefore, we would like to point out the 
principal discrepancies between our results and those of 
\cite{Sib01,Sib02b,Sib02c} which we were unable to trace back 
to model differences of the two calculations. 

\begin{figure}
\centerline{\psfig{figure=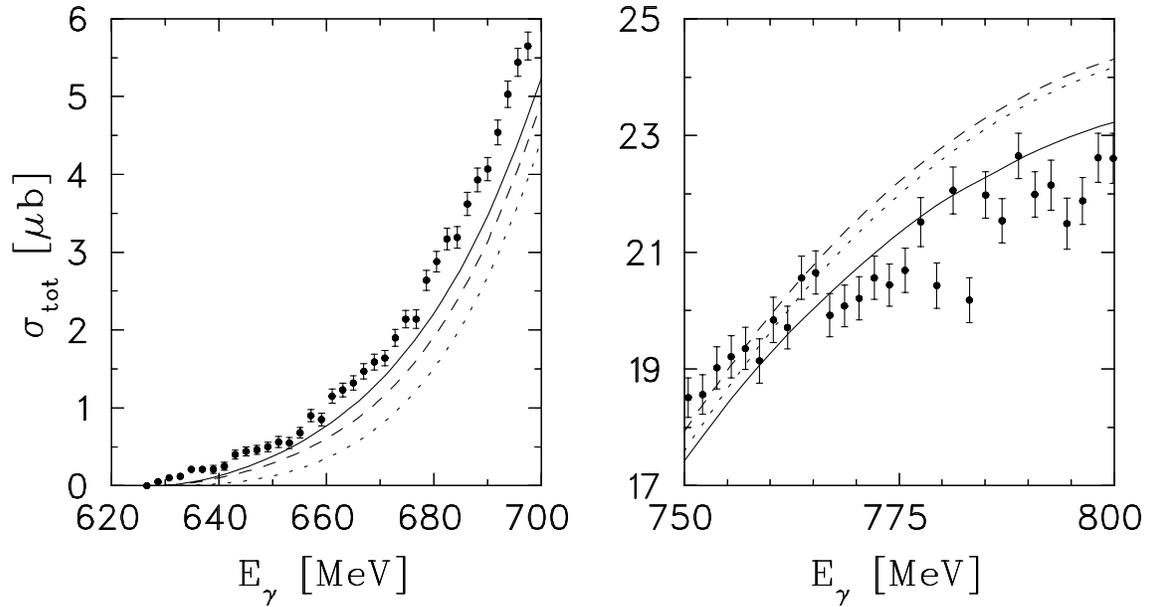,width=15cm,angle=0}}
\vspace{.5cm} 
\caption{
Total cross section of $\gamma d\to\eta np$ for two different regions of the 
photon energy. The dotted curve represents the IA. 
Successive addition of $NN$ and $\eta N$ rescattering in FOA 
is shown by the dashed and the solid curves, respectively. 
The $\gamma d\to\eta X$ data are from \protect\cite{Hej02}. 
} 
\label{fig4}
\end{figure}

\begin{figure}
\centerline{\psfig{figure=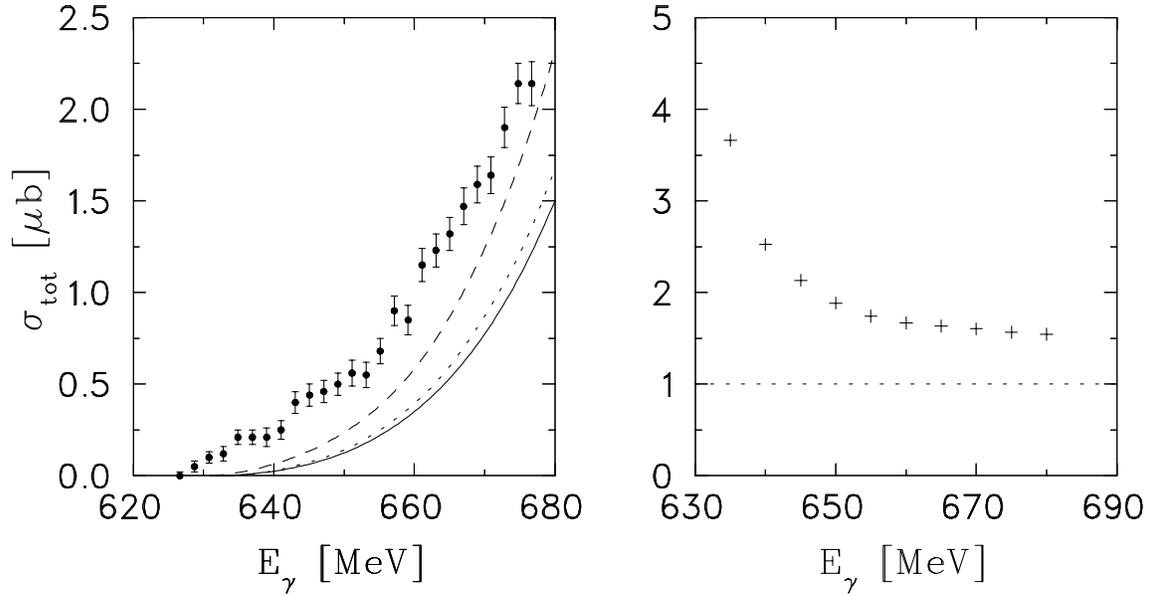,width=15cm,angle=0}}
\vspace{.5cm} 
\caption{Left panel:
comparison of the IA for the total cross section from 
\protect\cite{Sau96} (dotted curve), \protect\cite{Sib02c} (dashed curve), 
and for the present model (solid curve). 
The $\gamma d\to\eta X$ data are from \protect\cite{Hej02}. 
Right panel: ratio of the 
IA cross section of \protect\cite{Sib02c} to our calculation. 
} 
\label{fig5}
\end{figure}

To this end, we show in Fig.\,\ref{fig4} 
our first-order calculation of the total cross section, 
which reproduces our previous results obtained in \cite{FiAr97}. Small 
deviations are due to different parametrization of the elementary production 
operator and the $\eta N$ scattering amplitude.  
In the left panel of 
Fig.\,\ref{fig5} we compare our impulse approximation (IA) with the 
corresponding results of Fig.\,1 in \cite{Sib02c} and those presented 
in \cite{Sau96}. The right panel shows the ratio of the IA of \cite{Sib02c}
to our IA. One readily notices that our calculation is in 
reasonable agreement with that of \cite{Sau96} and the difference of our 
result
to the one of \cite{Sau96} indicates the model dependence to be expected 
from different parametrization of the elementary photoproduction amplitude 
and the deuteron wave function. But both IAs are far below the 
prediction of \cite{Sib02c}. 
For example, at the energy $E_\gamma = 635$ MeV the latter cross section 
is by about a factor 3.6 larger than ours (see 
right panel in Fig.\,\ref{fig5}). 
This disagreement is especially surprising since the IA is quite insensitive 
to the model ingredients. Namely, if one uses a realistic deuteron 
wave function in conjunction with an elementary production operator 
fitted to the single nucleon data, than the $\gamma d\to\eta np$ 
cross section is fixed almost unambiguously. A little freedom 
associated with the choice of the invariant energy $W_{\gamma N}$ 
of the active $\gamma N$ subsystem does not play a role at all, since in 
the present calculation and in \cite{Sib02c} 
this energy is taken according to the same prescription 
(compare the formulas (33) of \cite{FiAr97} and (2) of \cite{Sib02c}). 
In other words, a model dependence cannot explain
such a strong difference between the IA results. 

Turning now to the role of FSI we also find quite different effects. 
Namely in \cite{Sib02c} the FSI effect decreases rapidly above  
the threshold and vanishes completely at  
$E_\gamma$ = 680 MeV, so that above this point the total cross section is
determined exclusively by the IA, whereas according to our findings  
the FSI contribution does not exactly vanish at all energies   
examined here. In more detail, the enhancement of the cross section 
due to the $NN$ interaction is reduced asymptotically to about 
1.5 $\%$ at $E_\gamma = 780$ MeV (see Fig.\,\ref{fig4}).
For the $\eta N$ interaction we obtain an enhancement 
of about 20$\%$ at $E_\gamma = 635$ MeV.
This effect is visibly smaller than the 45 $\%$ 
given in Fig.\,6 of \cite{Sib02b} for the same energy \cite{foot1}.
This disagreement cannot be traced back to the strength of the $\eta N$
interaction because the $\eta N$ scattering length 
$a_{\eta N} = (0.42+i0.34)$ fm used in 
\cite{Sib02b} is even slightly smaller than our 
value given in (\ref{eq50}). 
With increasing energy the influence of the 
$\eta N$ FSI changes sign and results, according to our calculation, 
in a slight reduction 
of about 4 $\%$ of the cross section in the resonance peak 
(see right panel of Fig.\,\ref{fig4}). The latter  
effect must originate from a relatively strong absorption of $\eta$-mesons 
through the rescattering into pions, which is expected to be most 
pronounced in the 
resonance region. The same smearing of the resonance peak due to the strong
inelasticity of the $\eta N$ interaction is observed also 
in heavier nuclei \cite{Land96} where it manifests itself naturally much 
stronger. This energy dependence of the $\eta N$ FSI is  
not supported by the results presented in \cite{Sib02b,Sib02c} 
where rescattering effects do not play any role 
above $E_\gamma=680$ MeV. We would like to note that this  
discrepancy must be explained before any conclusion about the 
understanding of $\eta$ photoproduction on the deuteron is drawn. 


\section{The role of the $\eta N$ interaction}\label{sect4}

In this last section we address the question to what extent 
the $\eta N$ interaction can be ``extracted'' from 
incoherent $\eta$ photoproduction on the deuteron. 
The idea to obtain the information on the $\eta N$ low-energy 
scattering parameters from this reaction was explored in 
\cite{Sib02b,Sib02c}. In particular, it was found that 
the cross section is quite sensitive to the $\eta N$ 
interaction strength, making it possible to study $\eta N$ scattering by  
analyzing the observed single particle spectra or angular distributions. 

\begin{figure}
\centerline{\psfig{figure=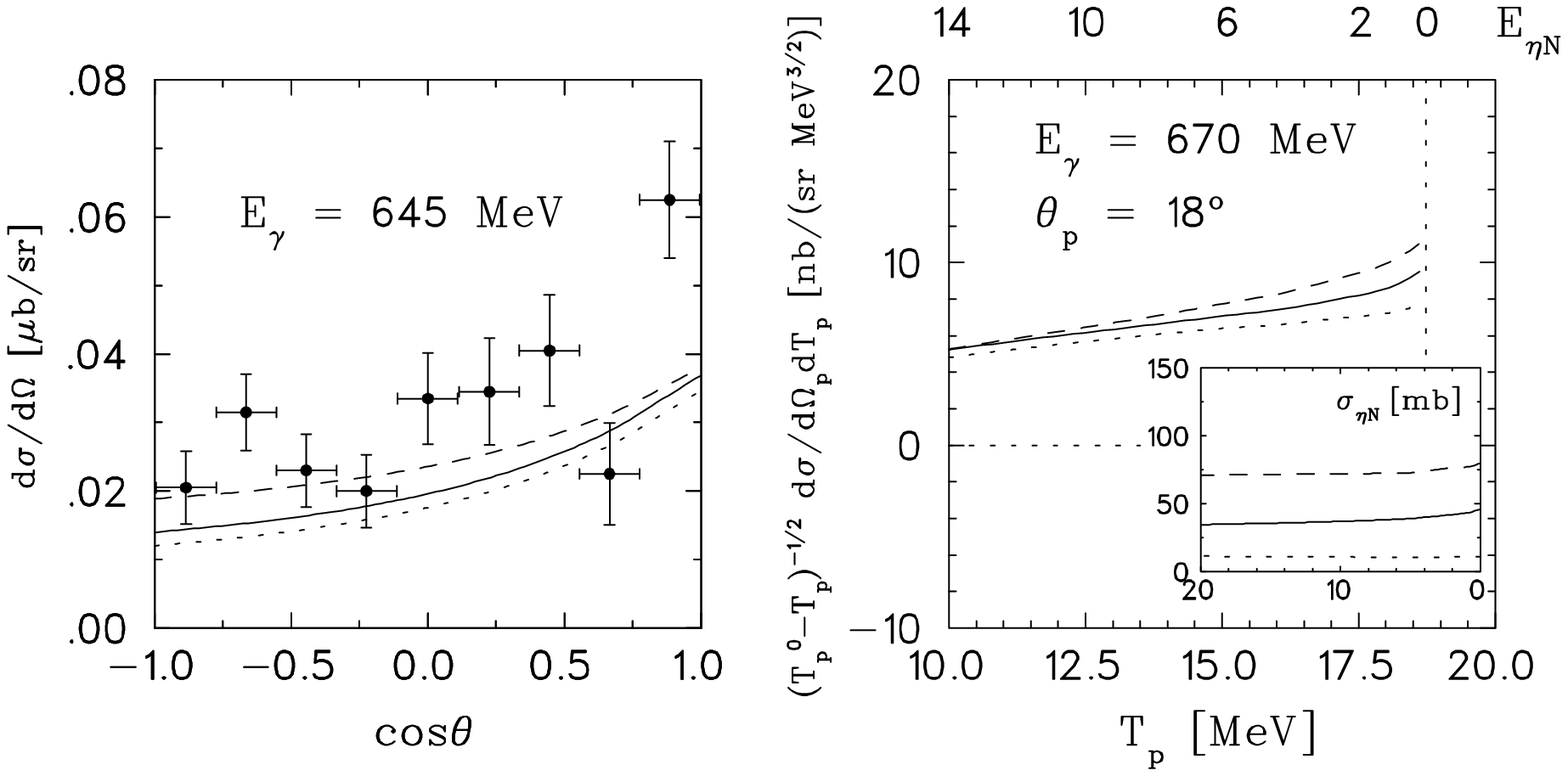,width=15cm,angle=0}}
\vspace{.5cm} 
\caption{Left panel:
The $\eta$ angular distribution calculated with different 
parametrization of the $\eta N$ scattering amplitude. 
Notation of the curves: dotted: $a_\eta N$ = (0.25+i0.16) fm;
full: $a_\eta N$ = (0.50+i0.32) fm; dashed: $a_\eta N$ = (0.75+i0.27) fm. 
The $\gamma d\to\eta X$ data are taken from \protect\cite{Hej02}. 
Right panel: Reduced proton 
energy spectrum, calculated at a fixed angle. Shown is the region in the 
vicinity of the maximum energy. The internal $\eta n$ kinetic energy 
$E_{\eta N}$ is indicated at the top x-axis. 
Insert: $\eta N$ elastic scattering cross section 
as a function of the energy $E_{\eta N}$. Notation of the curves as in the 
left panel.} 
\label{fig6}
\end{figure}

In our opinion, such a method very likely will meet serious difficulties,  
and we support our scepticism by several numerical results presented below. 
Firstly, and this is the crucial point for the following conclusions, 
we do not observe any strong sensitivity of the cross section to the 
$\eta N$ interaction strength as is claimed in \cite{Sib02b}. 
The effect on the $\eta$ angular distribution of varying the 
$\eta N$ parameters is demonstrated in the left panel of Fig.\,\ref{fig6}.  
The calculations 
are performed within the three-body approach as outlined above. 
In each case the e.m.\ vertex was adjusted to reproduce the 
elementary experimental cross section. Comparing the present result 
with Fig.\,4 of \cite{Sib02c} one 
sees that in our case the dependence on the $\eta N$ scattering length 
is much less pronounced. Thus, we have to conclude that the experimental 
discrimination between different $\eta N$ models would require extremely 
precise measurements. 

Furthermore, such a weak sensitivity makes the extraction of the 
$\eta N$ scattering parameters very model dependent. The reason has to do with 
the off-shell behavior of the $\gamma N\to\eta N$ amplitude. 
This is especially critical for 
the region below the free $\eta N$ threshold which comes into play, when 
contributions beyond the IA are considered. 
For example, the results presented above are 
obtained within the so-called ``spectator on-shell'' choice for the invariant  
energy $W_{\gamma N}$ of the $\gamma N$ system, which is natural for the 
nonrelativistic three-body theory. In this case, as one integrates over the 
spectator nucleon momentum in the deuteron, $W_{\gamma N}$ covers the range 
$[-\infty, W-M_N]$ where $W$ is the invariant energy of the $\eta NN$ 
system. Although the uncertainty of the $\gamma N\to\eta N$ subthreshold 
behavior is not very crucial for the incoherent reaction, it  
makes the method of precise determination of $\eta N$ parameters 
much more ambiguous, than was presented in \cite{Sib02b}. 
This difficulty is quite general when one tries to determine the contribution 
of an individual diagram to the whole amplitude. 

In this connection we would like to recall   
the Migdal-Watson theory \cite{Migd77,Wat52}
which makes it possible to study the two-body interaction 
without regard to the particular way of embedding this interaction 
into the reaction amplitude. According to this theory the low-energy 
parameters of two-body scattering can in principle 
be identified by fitting the energy spectrum of the third particle 
close to the maximum energy value. In order to illustrate 
the practical applicability of this method 
to the $\eta N$ interaction, we show in the right panel of 
Fig.\,\ref{fig6} the proton spectrum at a fixed
angle $\theta_p = 18^o$. The curves were obtained within the three-body 
approach using the same three different parametrizations of the 
$\eta N$ sector as in the left panel. Only the upper end of the spectrum, 
corresponding to low relative $\eta n$ energies, is shown, since 
only this part is relevant. According to the Migdal-Watson theory 
the behavior of the proton energy distribution in this region 
reflects the energy dependence of 
the elementary $\eta n$ cross section, disregarding the terms which 
depend weakly on the $\eta n$ relative energy. To eliminate the effect of the 
kinematical boundary, the spectrum is divided by $(T^0_p-T_p)^{1/2}$ 
where $T^0_p$ is the maximum proton energy. As one notes, 
also in this case the experimental discrimination between the $\eta N$
models seems to meet the same difficulties. Namely, the form of the
spectrum is not strongly affected by the $\eta N$ interaction. 
The reason is that the form of the 
$\eta N$ elastic scattering cross 
section, shown in the insert in the right panel,   
does not sizeably vary with the value of $a_{\eta N}$.


\section{Conclusion}\label{sect5}

The role of final state interaction in incoherent $\eta$ photoproduction
on a deuteron has been investigated 
using a three-body model for treating the interaction
in the final $\eta NN$ system. In contrast to our previous work \cite{FiAr02} 
the present results were obtained by using a realistic 
$NN$ interaction based on the separable representation of the Bonn potential. 
As a summary, we would like to draw the following conclusions:

(i) As may be expected, a realistic treatment of the $NN$ sector reduces 
the FSI effect compared to the use of a simple Yamaguchi potential. At 
the same time, the influence of higher-order rescattering in the final state 
remains essential and must be taken into account close to the threshold.

(ii) Our three-body calculation underestimate the data \cite{Hej02} 
slightly in the energy region up to $E_\gamma = $ 780 MeV. 
One open point, which remains to be investigated in this 
connection, is the short-range part of the $\eta NN$ wave function. The latter 
can be quite 
sensitive, e.g., to the contribution of the $\pi NN$ configuration 
which was omitted here. 

(iii) There exists a 
principal discrepancy between our results and the ones of Sibirtsev et al. 
\cite{Sib01,Sib02b,Sib02c}. 
Most disturbing is the fact that already the simple impulse approximation
of \cite{Sib01,Sib02b,Sib02c} where the model ambiguities should be small, 
exhibits a strong disagreement with other authors \cite{FiAr97,Sau96}. 
In this connection we would like to emphasize, that the statement, that 
the reaction $\gamma d\to\eta np$ is well understood within existing 
theoretical approaches is premature, even if a seemingly 
good description of the data in \cite{Sib02c} 
is achieved. Further theoretical work is certainly needed. 

(iv) We do not find a strong sensitivity of the $\gamma d\to\eta np$ cross 
section to the $\eta N$ interaction strength as was claimed in \cite{Sib02b}. 
Our calculation shows that even if the off-shell
uncertainty of the $\gamma N\to\eta N$ amplitude 
is disregarded, quite a weak dependence of the results on the $\eta N$
interaction parameters renders a precise determination of the $\eta N$ 
interaction practically impossible. 

\acknowledgments
The work was supported by the Deutsche Forschungsgemeinschaft (SFB 443).



\begin{table}
\renewcommand{\arraystretch}{1.7}
\caption{
Parameters of the $\eta N$ scattering matrix in eqs. 
(\protect\ref{eq10}), (\protect\ref{eq20}), and (\protect\ref{eq40}).}
\begin{tabular}{cccccc}
$a_{\eta N}$ [fm] & $g_\eta$ & $\beta_\eta$ [MeV] & $g_\pi/\sqrt{3}$  & 
$\beta_\pi$ [MeV] & $M_0$ [MeV] \\ \hline
0.25+i0.16 & 1.43 & 654.3 & 1.57 & 379.2 & 1563 \\
0.50+i0.32 & 2.00 & 694.6 & 1.45 & 404.5 & 1598 \\
0.75+i0.27 & 2.04 & 1282.6 & 0.55 & 888.0 & 1673 \\
\end{tabular}
\label{tab1}
\end{table}


\vspace{1cm}
\vspace{1cm}

\begin{thebibliography}{99}
\bibitem{foot1}
There is an inconsistency to the enhancement shown in 
Fig.\,5 of \protect\cite{Sib02b} where the $\eta N$ enhancement is
even larger, namely more than 100 $\%$ at $E_\gamma = 635$ MeV.
\end{thebibliography}
\end{document}